\documentstyle[aps,epsfig]{revtex}
\begin{document}
\draft
\title{Double giant dipole resonances in time-dependent density-matrix 
theory}
\author{Mitsuru Tohyama}
\address{Kyorin University School of Medicine, 
Mitaka, Tokyo 181-8611, Japan
\\
and
\\
Japan Atomic Energy Research Institute \\
Tokai-mura, Naka-gun, Ibaraki 319-1195, Japan}
\date{\today}
\maketitle
\begin{abstract}
The strength functions of the double giant dipole resonances (DGDR)
in $^{16}$O and $^{40}$Ca are calculated with the use of 
an extended version of the 
time-dependent Hartree-Fock theory known as the time-dependent 
density-matrix theory. The calculations are done in
a self-consistent manner, in which 
the same Skyrme force as that used for
a mean-field potential is used as an effective interaction for
a two-body correlation function.
It is found that the DGDR in $^{16}$O has a large width due to
the Landau damping, although the centroid energy of the strength
distribution is close to twice the energy of the giant dipole resonance
(GDR) calculated in RPA. The DGDR in $^{40}$Ca is found more harmonic than that in $^{16}$O:
the strength function of the DGDR in $^{40}$Ca is similar to
what is predicted from the strength function of the GDR in RPA.

\vspace{0.5cm}

\vspace{0.5cm}
\end{abstract}
%\pacs{PACS number(s): 21.60.Jz, 24.10.Cn, 24.30.Cz}
%\narrowtext

\vspace{0.5cm}
The double phonon states of giant resonances have become the subject
of a number of recent experimental and theoretical 
investigations \cite{Chomaz,Bertulani}. Microscopic calculations of 
the strength functions of double giant dipole 
resonances (DGDR)
have also been done 
based on the shell model \cite{Nishi1,Nishi2} and quasiparticle-phonon
models \cite{Bertulani,Dang}.
However, few microscopic studies have been reported, 
in which a single-particle basis and a residual interaction are
treated in such a self-consistent manner as used in 
random-phase-approximation (RPA)
calculations
for giant resonances \cite{Tsai}.
We have recently proposed a self-consistent approach \cite{T98} based on an
extended version of the time-dependent Hartree-Fock theory (TDHF) known
as the  time-dependent density-matrix theory (TDDM) \cite{GT1},
in which the same Skyrme force as that used for the 
calculation of a mean-field potential is used as a residual interaction
for a two-body correlation function.
We applied the model to the double giant quadrupole resonances (DGQR) in $^{16}$O 
and $^{40}$Ca \cite{T99} and showed that the
DGQR's in these nuclei have strong 
harmonic properties. The aim of this paper is to report the results
of the application of the TDDM approach to the DGDR's
in $^{16}$O and $^{40}$Ca. 

The formulation of TDDM is based on the truncation of the hierarchy
of reduced density matrices, in which   
genuine correlated parts in a three-body density matrix
and higher reduced density matrices are neglected
\cite{WC}. 
The TDDM equations thus determine the time evolution of 
a one-body density matrix $\rho$ and
a two-body correlation function $C_{2}$ defined by 
$C_{2} =\rho_{2} - A[\rho\rho]$, where $A[\rho\rho]$
is the antisymmetrized product of the one-body density 
matrices and $\rho_{2}$ is a two-body density matrix.
In TDDM, further truncation is made by expanding
$\rho$ and $C_{2}$ with a finite number of single-particle states
$\{\psi_{\alpha}\}$ as
\begin{eqnarray}
\rho(11',t)&=&\sum_{\alpha\alpha'}n_{\alpha\alpha'}(t)\psi_{\alpha}(1,t)
\psi_{\alpha'}^{*}(1',t), 
\end{eqnarray}
\begin{eqnarray}
C_{2}(121'2',t)
=\sum_{\alpha\beta\alpha'\beta'}C_{\alpha\beta\alpha'\beta'}(t)
\psi_{\alpha}(1,t)\psi_{\beta}(2,t)
\psi_{\alpha'}^{*}(1',t)\psi_{\beta'}^{*}(2',t),
\end{eqnarray}
where the numbers denote space, spin and isospin coordinates.
The time evolution of $\rho$ and $C_{2}$ is determined by the following
three coupled equations \cite{GT1}:
\begin{eqnarray}
i\hbar\frac{\partial}{\partial t}\psi_{\alpha}(1,t)=h(1,t)
\psi_{\alpha}(1,t),
\end{eqnarray}
\begin{eqnarray}
i\hbar \dot{n}_{\alpha\alpha'}=\sum_{\beta\gamma\delta}
[\langle\alpha\beta|v|\gamma\delta\rangle C_{\gamma\delta\alpha'\beta}
-C_{\alpha\beta\gamma\delta}\langle\gamma\delta|v|\alpha'\beta\rangle],
\end{eqnarray}
\begin{eqnarray}
i\hbar\dot{C}_{\alpha\beta\alpha'\beta'}=B_{\alpha\beta\alpha'\beta'}
+P_{\alpha\beta\alpha'\beta'}+H_{\alpha\beta\alpha'\beta'},
\end{eqnarray}
where
$h(1,t)$ is the mean-field hamiltonian and $v$ the residual interaction.
The term $B_{\alpha\beta\alpha'\beta'}$ on the right-hand side of Eq.(5)
represents the Born terms (the first-order terms of $v$). The terms 
$P_{\alpha\beta\alpha'\beta'}$ and $H_{\alpha\beta\alpha'\beta'}$ in Eq.(5)
contain $C_{\alpha\beta\alpha'\beta'}$ and 
represent higher-order particle-particle (and hole-hole) 
and particle-hole type correlations,
respectively. 
Thus full two-body correlations including
those induced by the Pauli exclusion principle
are taken into account in the equation of motion for $C_{\alpha\beta\alpha'\beta'}$.
The explicit expressions for $B_{\alpha\beta\alpha'\beta'}$,
$P_{\alpha\beta\alpha'\beta'}$ and $H_{\alpha\beta\alpha'\beta'}$ are
given in Ref.\cite{GT1}. The small amplitude limit of
TDDM was investigated \cite{TG1} and it was shown that
TDDM can be reduced to the second RPA \cite{Saw}-\cite{Drodz}
in such a limit. 
The number of two-body matrices treated in TDDM grows very rapidly with 
increasing mass number,
restricting the application of TDDM to light nuclei for
the present. To solve the coupled equations, 
we use the Skyrme interaction of the form \cite{Vaut}
\begin{eqnarray}
v({\bf r}-{\bf r'})&=&t_0(1+x_0P_{\sigma})\delta^3 ({\bf r}-{\bf r'})+
\frac{1}{2}t_1\{k'^2\delta^3 ({\bf r}-{\bf r'})+\delta^3 ({\bf r}-{\bf r'})k^2\} \nonumber \\
&+&t_2{\bf k'}\cdot\delta^3 ({\bf r}-{\bf r'}){\bf k}
+\frac{1}{2}t_3\rho\left(\frac{{\bf r}+{\bf r'}}{2}\right)
\delta^3 ({\bf r}-{\bf r'}),
\end{eqnarray}
where ${\bf k}=(\nabla_{\bf r} - \nabla_{\bf r'})/2i$ acts to the right and ${\bf k'}=(\nabla_{\bf r'} - \nabla_{\bf r})/2i$ acts to the left.  
The factor 1/2 on the density dependent term contains 
the contribution of a rearrangement effect \cite{Vaut}.
We use the parameter set of the Skyrme III force (SKIII) \cite{Bei}.
The spin-orbit force is neglected.
We assume that the motion of the
DGDR 
is generated by a two-body operator $\hat{D}^2$:
\begin{eqnarray}
|\Psi(t=0)\rangle=e^{ik\hat{D}^2}|\Phi_0\rangle,
\end{eqnarray}
where $\hat{D}$ is a one-body dipole operator and $|\Phi_0\rangle$ the ground-state
wave function. The initial conditions for solving the coupled equations 
Eqs.(3)-(5) are determined with the use of the above wave function.
We evaluate the 
initial values of $C_{\alpha\beta\alpha'\beta'}$
\begin{eqnarray}
C_{\alpha\beta\alpha'\beta'}(t=0)=
\langle\Psi(t=0)|a^+_{\alpha'}a^+_{\beta'}a_{\beta}a_{\alpha}
|\Psi(t=0)\rangle,
\end{eqnarray}
assuming that $|\Phi_0\rangle$ is
the Hartree-Fock (HF) ground-state wave function. 
At first order of $k$,
the initial condition for 
$C_{\alpha\beta\alpha'\beta'}$ becomes 
\begin{eqnarray}
C_{\mu\nu\rho\sigma}&=&\langle\Psi|a^+_{\rho}a^+_{\sigma}a_{\nu}a_{\mu}|\Psi\rangle
\nonumber \\
&=&2ik\{\langle\mu|D|\rho\rangle\langle\nu|D|\sigma\rangle
-\langle\mu|D|\sigma\rangle\langle\nu|D|\rho\rangle\}\\
C_{\rho\sigma\mu\nu}&=&\langle\Psi|a^+_{\mu}a^+_{\nu}a_{\sigma}a_{\rho}|\Psi\rangle
\nonumber \\
&=&-2ik\{\langle\rho|D|\mu\rangle\langle\sigma|D|\nu\rangle
-\langle\rho|D|\nu\rangle\langle\sigma|D|\mu\rangle\},
\end{eqnarray}
where $\rho$ and $\sigma$ refer to unoccupied single-particle 
states, and $\mu$
and $\nu$ refer to occupied ones. We choose the dipole operator 
in the above equations
such as $D=\tau_{z} z$.
Other elements of the initial $C_{\alpha\beta\alpha'\beta'}$
vanish at first order of $k$.
Similarly, non-varnishing 
initial values of $n_{\alpha\alpha'}$ become 
\begin{eqnarray}
n_{\mu\rho}&=&\langle\Psi|a^+_{\rho}a_{\mu}|\Psi\rangle
\nonumber \\
&=&2ik\sum_{\nu}\langle\mu|D|\nu\rangle\langle\nu|D|\rho\rangle\\
n_{\rho\mu}&=&\langle\Psi|a^+_{\mu}a_{\rho}|\Psi\rangle
\nonumber \\
&=&-2ik\sum_{\nu}\langle\rho|D|\nu\rangle\langle\nu|D|\mu\rangle .
\end{eqnarray}
For the initial $\psi_{\alpha}$'s 
we use the HF single-particle wave functions.
The strength function of the DGDR, defined by
\begin{eqnarray}
S_2(E)=\sum_{n}|\langle\Phi_n|\hat{D}^2|\Phi_0\rangle|^{2}\delta (E-E_{n}),
\end{eqnarray}
is given by the Fourier transform of
the time-dependent two-body dipole moment $D_2(t)$ as
\begin{eqnarray}
S_2(E)=\frac{1}{\pi k\hbar}\int_{0}^{\infty}D_2(t)\sin\frac{Et}{\hbar}dt,
\end{eqnarray}
where $D_2$ is given by
\begin{eqnarray}
D_2(t)&=&\langle\Psi(t)|\hat{D}^2|\Psi(t)\rangle \nonumber \\
&=&\sum_{\alpha\alpha'}\langle\alpha|D^2|\alpha'\rangle
n_{\alpha'\alpha} +
\sum_{\alpha\beta\alpha'\beta'}\langle\alpha|D
|\alpha'\rangle\langle\beta|D|\beta'\rangle
\{A[n_{\alpha'\alpha}n_{\beta'\beta}]+C_{\alpha'\beta'\alpha\beta}\}.
\end{eqnarray}
The terms without $C_{\alpha\beta\alpha'\beta'}$ in the above
equation have
negligible contribution to the Fourier transformation in Eq.(14).
The $k$ dependence of $S_2(E)$ thus obtained is negligible as long as $k$ is
sufficiently small.
The energy-weighted sum rule (EWSR) for the DGDR is given as
\begin{eqnarray}
\int_0^{\infty} ES_2(E)dE &=& \frac{1}{2}
\langle\Phi_0|[\hat{D}^2,[H,\hat{D}^2]|\Phi_0\rangle \nonumber \\
&=& \frac{2\hbar^2}{m}\langle\Phi_0|\hat{D}^2|\Phi_0\rangle
+4(t_1+t_2)\langle\Phi_0|\hat{R}\hat{D}^2|\Phi_0\rangle,
\end{eqnarray}
where $H$ is the total hamiltonian and $m$ the nucleon mass.
The second term on the right-hand side of Eq.(16) is due to the momentum
dependence of the Skyrme force and 
$\hat{R}$ is the following two-body operator
\begin{eqnarray}
\hat{R}=\sum_{i\in p, j\in n} \delta^3 ({\bf r}_i-{\bf r}_j).
\end{eqnarray}
The EWSR value is evaluated with the use of
the HF wave function for $|\Phi_0\rangle$. The second term on
the right-hand side of Eq.(16) has a contribution of about 30\% to the
total EWSR value.

To solve the coupled equations Eqs.(3)-(5), we use a minimum number of
single-particle states: the $1s, 1p,
2s$ and $1d$ single-particle orbits for $^{16}$O and the
$1s, 1p, 2s,1d,
2p$ and $1f$ orbits for $^{40}$Ca. 
To check the validity of such truncation of the 
single-particle space, we performed RPA calculations for 
the GDR strength functions in $^{16}$O and $^{40}$Ca
using the time-dependent RPA equations \cite{T99} in the same truncated
space.
The obtained results were compared with those of the 
TDHF calculations which correspond to continuum 
RPA calculations \cite{Tsai}.
The fractions of the EWSR values depleted in the 
energy interval $0-40$MeV were turned out to be
90\% in $^{16}$O and 93\% in $^{40}$Ca, respectively. These
values are
sufficiently large and comparable with the TDHF values which are
close to 100\% \cite{T982}. However,
the excitation energies of the GDR's in RPA were slightly larger
than those in TDHF. To adjust the excitation energies of the GDR's
to the TDHF values, we
reduced the parameter $x_0$ of the spin-dependent term of the 
Skyrme force Eq.(6). The obtained
value of $x_0$ was 0.3 instead of the original value of 0.45.
We use this reduced value of $x_0$ in the following calculations.
The spin-dependent term of the Skyrme force 
has a negligible contribution to the mean-field
potential in spin-isospin symmetric nuclei like $^{16}$O and $^{40}$Ca
considered here \cite{Vaut} and, therefore, the single-particle wave 
functions are not affected by the reduction of the strength of
the spin-dependent term. 
The integration in Eq.(14)
is performed for a finite time interval of $1.5\sim 2\times10^{-21}$s.
As a result
$S_2(E)$ has small fluctuations.
To reduce the fluctuations in $S_2(E)$, we multiply
$D_2(t)$ 
by a damping factor
$e^{-\Gamma t/2}$
before performing the time integration. 
This corresponds to smoothing the strength function with a
width $\Gamma$. We use $\Gamma=1$MeV.
Other calculational details are explained 
in our previous publications 
\cite{T99,T}.

\begin{figure}
\epsfig{file=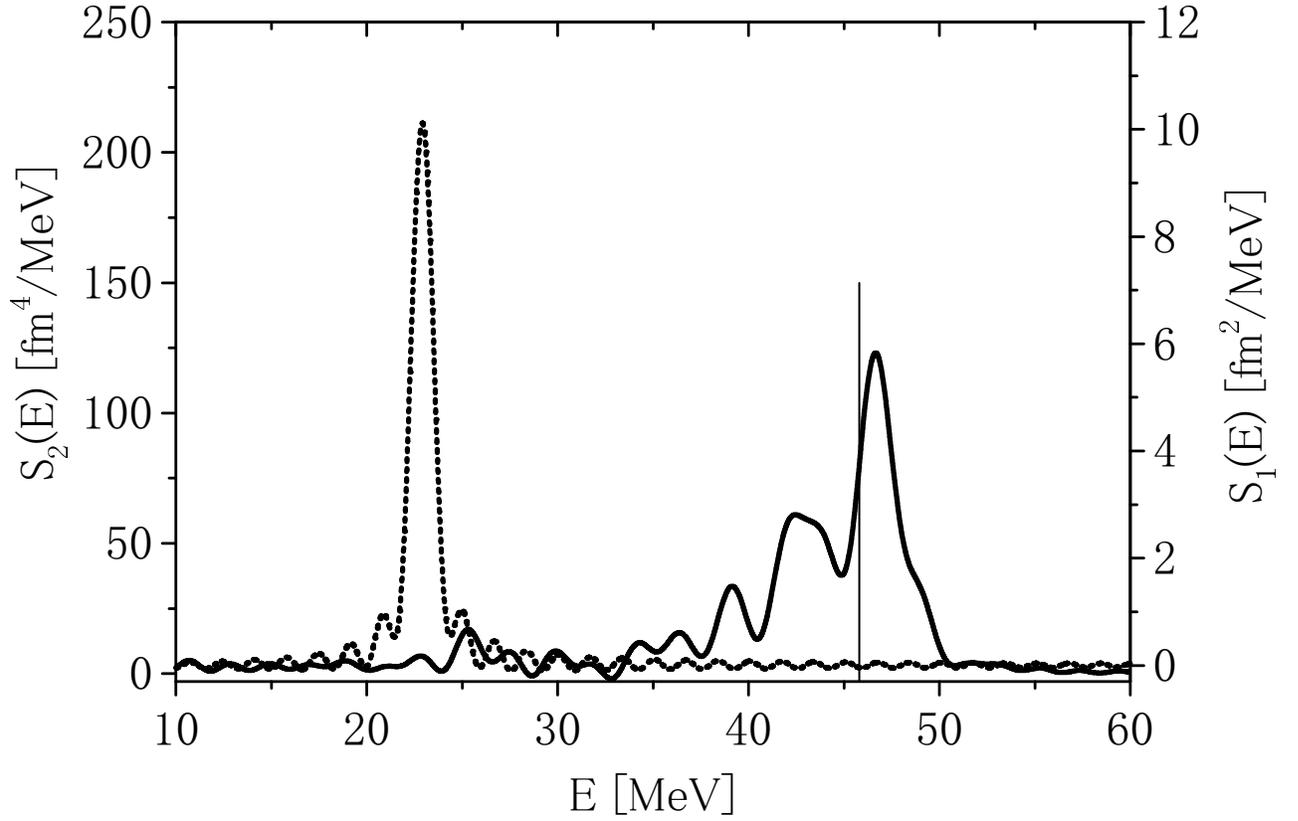,width=1\textwidth}
\caption{Strength function $S_2(E)$ of the DGDR in $^{16}$O calculated in TDDM 
(thick solid line).
The dotted line denotes the strength function $S_1(E)$ 
of the GDR in RPA and
the thin vertical bar at $E=45.8$MeV indicates
the location of the DGDR predicted from the GDR in RPA (in
arbitrary units).
}
\label{ODGDR}
\end{figure}

The strength distribution of the DGQR in $^{16}$O
calculated in TDDM 
is shown in Fig.\ref{ODGDR} (thick solid line).
The bump seen around $E=45$MeV 
corresponds to the DGDR. The
strength function $S_1(E)$ of the GDR obtained from
the time-dependent  
RPA calculation is also shown in Fig.\ref{ODGDR} (dotted line). 
The width of the GDR is small and
nearly equal to the width due to the smoothing and the finite time
integration as explained above.
The thin vertical bar at $E=45.8$MeV
indicates the location of the DGDR strength predicted from the GDR
shown in Fig.\ref{ODGDR}. 
The fraction of the EWSR value of the DGDR depleted in the 
energy interval 10--60MeV is 82\%. The centroid energy of the DGDR
strength distribution between 35MeV and 55MeV is 44.6MeV. 
The energy difference 
$\Delta E=45.8$MeV$-44.6$MeV$=1.2$MeV is small but 
slightly larger than 
the value
$\Delta E\approx0.8$MeV obtained from the work done by de 
Souza Cruz and
Weiss \cite{Cruz} using the generator
coordinate method and  
the value $\Delta E<0.7$MeV obtained from the formula 
given
by Bertsch and Feldmeire \cite{Bertsch}.
Though the centroid energy of the DGDR is close to twice the GDR energy, 
the DGDR in $^{16}$O has a large width due to the Landau
damping.

\begin{figure}
\epsfig{file=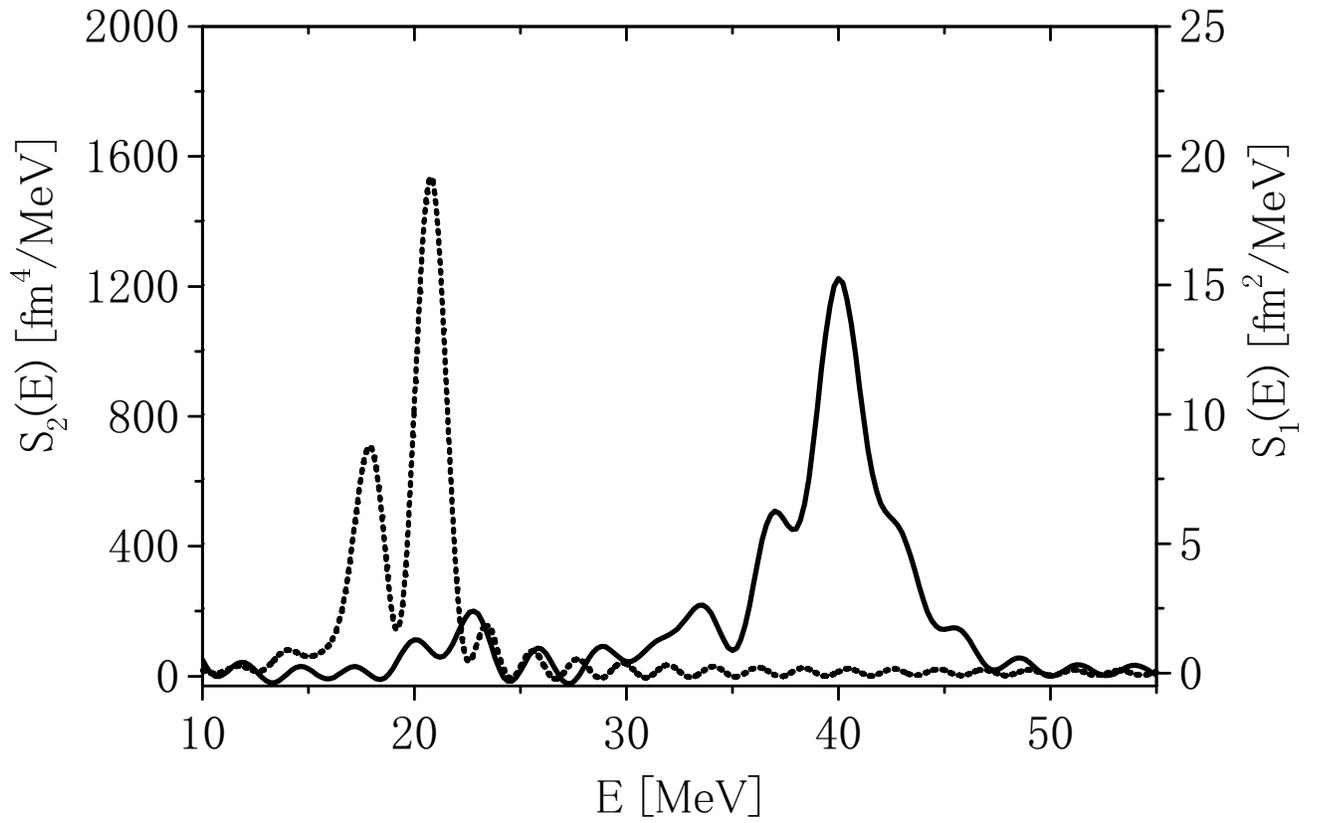,width=1\textwidth}
\caption{Strength function $S_2(E)$ of the DGDR in $^{40}$Ca calculated in TDDM (solid line).
The dotted line denotes the strength function $S_1(E)$ of the GDR in RPA.
}
\label{CaDGDR1}
\end{figure}

\begin{figure}
\epsfig{file=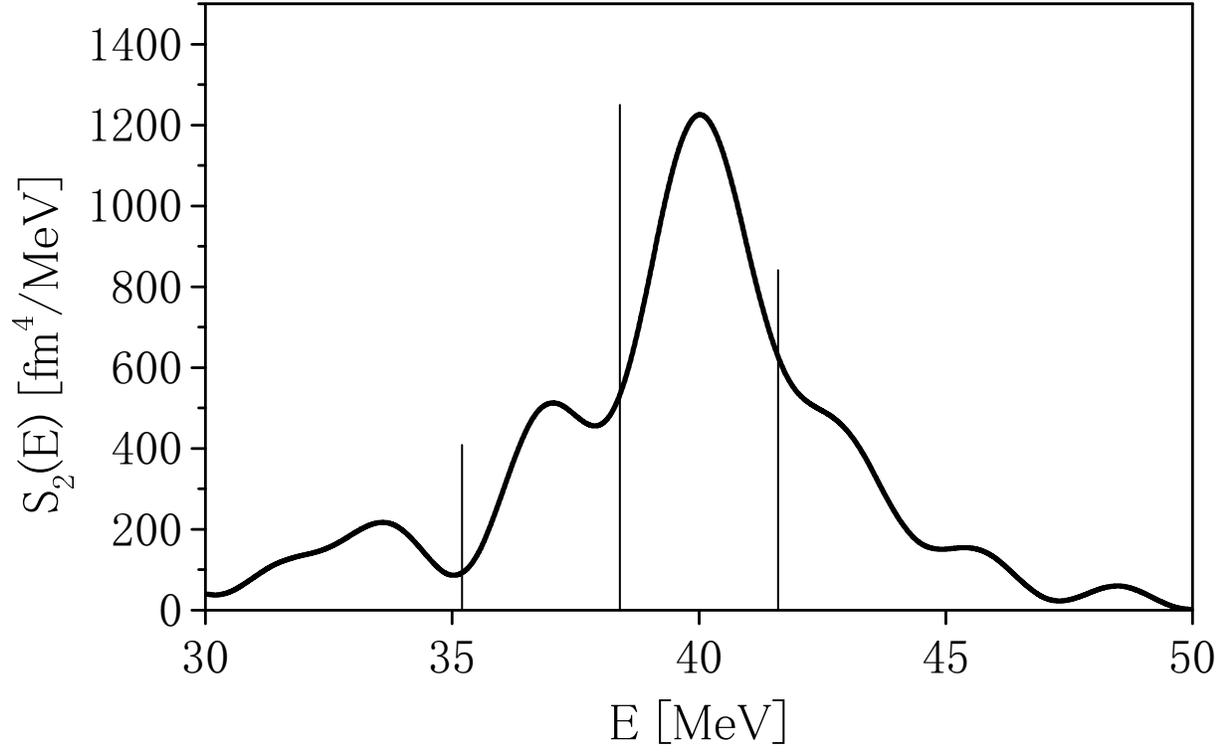,width=1\textwidth}
\caption{Strength function $S_2(E)$ of the DGDR in $^{40}$Ca calculated in TDDM 
(thick solid line)
is compared with what is predicted from the strength function of the GDR shown in 
Fig.\ref{CaDGDR1} (thin vertical bars in arbitrary units).}
\label{CaDGDR2}
\end{figure}

The strength distribution of the DGQR in $^{40}$Ca calculated in TDDM 
is shown in Fig.\ref{CaDGDR1} (solid line).
The bump seen around $E=40$MeV 
corresponds to the DGDR.
The strength function $S_1(E)$ of the GDR in $^{40}$Ca calculated in 
the time-dependent RPA is also shown in Fig.\ref{CaDGDR1} (dotted line). 
The GDR strength is split into two peaks in the 
case of $^{40}$Ca. A similar split is seen in the TDHF calculation for $^{40}$Ca
\cite{T982}.
The fraction of the EWSR value of the DGDR depleted in the 
energy interval 10--60MeV is 88\%. 
In Fig.\ref{CaDGDR2} the strength function of the DGDR (thick solid line) 
is compared with what is expected from 
that of the GDR shown in Fig.\ref{CaDGDR1}: 
The thin vertical bars in Fig.\ref{CaDGDR2}
indicate the locations and relative strengths of the DGDR 
predicted from the GDR in RPA, assuming that the GDR consists of the two
discrete components.
The centroid of the DGDR strength distribution in the
energy range 30--50MeV is 39.6MeV, while that of the 
GDR prediction is 39.0MeV.
The energy difference $\Delta E =0.6$MeV may be larger than 0.2MeV
obtained from Refs.\cite{Cruz,Bertsch} but is smaller than
1.0MeV given by the shell model calculation for $^{40}$Ca \cite{Nishi1}. 
Though the main peak in TDDM located at 40MeV has more strengths than 
that predicted
from the RPA calculation, the DGDR seems to have weaker Landau damping in $^{40}$Ca
than in $^{16}$O 
and the shape of 
the strength distribution is similar to
what is expected from the GDR in RPA. This means that the dipole mode 
becomes fairly harmonic in the mass region of $^{40}$Ca. 
A similar conclusion was obtained by de Souza Cruz and
Weiss \cite{Cruz} comparing the DGDR in $^{16}$O with that in $^{40}$Ca.
The small anharmonicities of the DGDR in Ca have also 
been pointed out by Catara et al.\cite{Catara} using a boson expansion 
approach and the anharmonicities of the DGDR's in heavier nuclei have recently
been studied by several groups \cite{Chomaz,Bertulani,Nishi2,Dang,Lanza}.  

In summary, the strength functions of the DGDR's in $^{16}$O and $^{40}$Ca
were calculated in TDDM in a self-consistent manner, in which the 
same Skyrme force as that used for the calculation of
the mean-field potential was used 
for the two-body correlation function.
It was pointed out that
the strength function of the DGDR is obtained from the Fourier
transform of the time-dependent two-body dipole moment.
It was found that in both nuclei 
the excitation energies of the DGDR's are very close to
twice those of the GDR's. It was also found that the
DGDR in $^{16}$O has a large width due to 
the Landau damping, indicating the anharmonicity of the dipole
mode in $^{16}$O.

\newpage
\newpage
\newpage
\newpage
\end{document}